# Les connaissances de la toile

Serge Abiteboul

INRIA, ENS Cachan, Conseil national du numérique & Académie des sciences

Les données tiennent depuis toujours une place essentielle dans le développement de l'informatique. Depuis les années 60, les logiciels de bases de données se sont imposés pour permettre le partage des données à l'intérieur d'une entreprise ou d'une organisation. Ces données qui étaient isolées dans des centres de calcul sont devenues accessibles partout dans le monde avec l'arrivée d'Internet, le réseau des *réseaux de machines*. Et puis est arrivé, le Web – la Toile – le *réseau de contenus*. On a vite réalisé que cela rendait possible le rêve de la connaissance et la culture accessibles par tous. Evidemment, ce rêve est « à des détails près » comme l'e-exclusion ou la propriété privé de certains contenus. L'étape suivante a été le développement des réseaux sociaux – des *réseaux d'individus* – basés sur le partage d'informations personnelles, la communication, la création de communauté, l'internaute passant de simple consommateur à producteur d'information.

Nous baignons aujourd'hui dans un monde numérique. Pour donner jusque quelques chiffres. Nous sommes entourés de milliards d'objets communicants. La toile en 2008 comptait déjà plus de 1000 milliards de pages et chaque mois, les internautes réalisaient des dizaines de milliards de recherches Web. Et surtout, on pense que le monde numérique double tous les 18 mois. Le trafic sur Internet est déjà chaque année supérieur à tout ce que nous pourrions stocker en utilisant tous les supports, tous les disques disponibles.

Au cœur de ce monde de la Toile, on trouve l'information principalement sous forme textuelle. Les individus aiment écrire, lire, dire, écouter du texte alors que les machines préfèrent des connaissances formatées. Pour que le Web puisse être pleinement utilisé par des machines à notre service, il fait passer de l'information aux connaissances. On commence à voir de telles connaissances sous la forme d'annotations de ressources du Web ; c'est le Web sémantique. On voit aussi arriver des bases de connaissances, des « ontologies », dans des domaines spécifiques ou même de manière plus générale. Dans cette quête de connaissances, la grande difficulté est d'obtenir des connaissances. Elles peuvent être introduites « manuellement », notamment en utilisant des éditeurs de connaissance. Ce modèle fonctionne parfaitement dans des cadres industriels ou scientifiques. En particulier, on considère que la constitution d'une base de connaissances ou de données est une tâche noble pour un scientifique, qui peut être aussi valorisante que l'écriture d'un article scientifique. Mais c'est moins simple dans le cadre général du Web avec des internautes qui, s'ils aiment écrire, n'adorent pas éditer des ontologies.

Il faut donc s'appuyer sur l'extraction automatique de connaissances de la Toile. C'est par exemple, ce qui a été réalisé à partir de l'encyclopédie Wikipedia pour obtenir l'ontologie Yago, ou dans des travaux sur l'alignement d'ontologies où j'ai pu participer avec le système Paris. La difficulté dans ces réalisations est la nature imprécise de l'information de nature linguistique.

On pourra mentionner une foule d'approches pour dégager collectivement des connaissances. Les internautes peuvent être amenés à donner des avis. C'est ce que fait un webmaître quand il met un hyperlien vers un autre site, ou un client de eBay qui note le service qu'il vient d'utiliser. Les systèmes sont de plus en plus doués pour recommander des produits comme dans Amazone ou des partenaires dans Meetic, en se basant sur des analyses statistiques de corrélation. Les internautes peuvent même proposer plus activement leurs compétences pour, par exemple, du co-développement ou pour la résolution de problèmes en crowd sourcing.

Les algorithmes tiennent explicitement une place de plus en plus importante dans nos vies. (Ils étaient présents implicitement : par exemple, en nous habillant nous suivons un algorithme appris très jeune.) Nous vivons de plus en plus dans un monde entourés de systèmes qui traitent de l'information numérique pour nous. Surtout, nous passons d'un monde fermé et précis, à un monde ouvert, imprécis, parfois incohérent. Qu'on le veuille ou pas, il va falloir vivre dans ce monde là.

En conclusion, cela nous ramène à ces mots de Manuel Castells : « La fracture numérique ne sépare pas tant ceux qui ont un accès à l'internet de ceux qui n'en ont pas, mais ceux qui savent quoi en faire culturellement de ceux pour qui ce n'est qu'un écran d'annonces accompagné de passe-temps ludiques. » Pour nous, cela soulève un sujet essentiel, celui de l'enseignement de l'informatique. Un tel enseignement est indispensable pour comprendre le monde qui nous entoure. L'école, le collège, et le lycée actuels ont été définis pour expliquer un monde qui n'est plus. L'enseignement de l'informatique est indispensable pour mieux utiliser les systèmes informatiques dans lesquels nous baignons, pour être maitre et pas esclave de ces nouvelles technologies, pour construire le monde de demain. Comme le propose le rapport récent de l'académie des sciences sur l'enseignement de l'informatique, il faut enseigner l'informatique pour tous les élèves à partir de l'école primaire et jusqu'au lycée et l'université.